\documentclass{rnaastex}

\begin{document}

\title{An extremely red and two other nearby L dwarf candidates
previously overlooked in 2MASS, WISE, and other surveys}

\correspondingauthor{Ralf-Dieter Scholz}
\email{rdscholz@aip.de, cbell@aip.de}

\author[0000-0002-0894-9187]{Ralf-Dieter Scholz}
\affiliation{Leibniz Institute for Astrophysics Potsdam (AIP),\\ 
An der Sternwarte 16, 14482 Potsdam, Germany}

\author[0000-0003-0642-6558]{Cameron P.M. Bell}
\affiliation{Leibniz Institute for Astrophysics Potsdam (AIP),\\
An der Sternwarte 16, 14482 Potsdam, Germany}


\section{}

The surprisingly uneven distribution of the nearest ($d < 6.5$\,pc)
known brown dwarfs \citep{2016A&A...589A..26B} suggests that there 
exist remaining undiscovered objects in this volume. 
We therefore extended previous
combined color/proper motion searches \citep{2011A&A...532L...5S,
2013A&A...557A..43B, 2014A&A...561A.113S, 2014A&A...567A..43S} to 
identify substellar neighbors among bright objects ($W2 < 13$\,mag) 
from the all-sky catalogue of the Wide-field Infrared Survey Explorer 
(WISE) \citep{2010AJ....140.1868W}. Compared to these previous searches,
we only slightly changed and complemented the WISE color and magnitude 
selection criteria, but included extended WISE sources and also looked
closer to the crowded Galactic plane. As in earlier studies, 
we expected that nearby 
L and T dwarf candidates were also observed by 
the Two Micron All Sky Survey (2MASS) \citep{2006AJ....131.1163S} but
have significant proper motions ($>100$\,mas/yr).

We cross-matched $\approx$15000 candidates for known counterparts 
in both SIMBAD and the 
compilation of 1886 known LTY dwarfs \citep{2017MNRAS.469..401S}.
Proper motions of $\approx$100 new candidates were visually detected on 
2MASS and WISE finder charts. For L dwarf candidates with $J-K_s>1.7$\,mag,
proper motions were then measured based on the highest-quality 
positions from all available WISE epochs (NEOWISE), 2MASS,  
the DEep Near-Infrared Survey (DENIS) \citep{1997Msngr..87...27E},
SuperCOSMOS Sky Surveys (SSS) \citep{2001MNRAS.326.1279H},
VISTA hemisphere survey (VHS) \citep{2013Msngr.154...35M},
INT photometric H$\alpha$ survey (IPHAS) \citep{2014MNRAS.444.3230B},
Panoramic Survey Telescope And Rapid Response System (PS1)
\citep{2017yCat.2349....0C}, and
the first Gaia data release \citep{2016A&A...595A...2G}.
Three new L dwarf candidates
(Table \ref{tab:1}) represent interesting bright
targets for various follow-up observations:

\textbf{2MASS~J07555430-3259589}, an extremely 
red ($J-K_s=2.72, J-W2=4.38, W1-W2=0.61$\,mag) L dwarf candidate, 
resembles one of the reddest young moving group (YMG) member
candidates (WISEA~090258.99+670833.1; L7) recently found
in a dedicated search \citep{2017AJ....153..196S}. Among the
few previously known peculiar (red) late-L dwarfs 
\citep{2017MNRAS.469..401S} we identified WISEP~J004701.06+680352.1
(L7.5p red) as a suitable comparison object with similar colors
and a trigonometric parallax \citep{2016ApJS..225...10F}. 
From this comparison, we classified
2MASS~J07555430-3259589 as $\approx$L7.5p and estimated a
photometric distance of $\approx$16\,pc. Using its distance,
position, 2MASS-WISE proper motion (PS1 and Gaia detected 
only a faint background object separated by $\approx$2\arcsec),
and assuming a radial velocity $|RV|<10$\,km/s, we find a 
Carina-Near YMG membership probability of 
$\approx$90\% \citep{2018arXiv180109051G}. 

\textbf{2MASS~J07414279-0506464} appears merged with a relatively bright 
background star in all WISE exposures. After excluding these and the
available DENIS and PS1 epochs, the proper motion fit was improved
significantly. The source was resolved in Gaia DR1 as a close binary
(separation 0.3\arcsec). The 2MASS and PS1 colors are typical of 
L5 dwarfs \citep{2018ApJS..234....1B}. Assuming an equal-mass binary
and using the mean absolute $J$ magnitude 
of single L5 dwarfs \citep{2012ApJS..201...19D},
we estimated a photometric distance of $\approx$19\,pc.

\textbf{2MASS~J19251275+0700362} was also merged with a background 
object in WISE images, and only one of two PS1 positions fitted
well in the proper motion solution. 2MASS and PS1 colors are fully
consistent with an L7 dwarf \citep{2018ApJS..234....1B}. The
$J$ magnitude equals the mean absolute magnitude of L7 dwarfs
\citep{2012ApJS..201...19D}. Therefore, this turned out to be
the nearest of our L dwarf candidates with a photometric 
distance of $\approx$10\,pc.

\begin{deluxetable}{lrrr}
\tablecaption{Astrometry, photometry, and estimated spectral types and distances of three new nearby L dwarf candidates.\label{tab:1}}
\tablehead{
\colhead{Parameter} & \colhead{2MASS~J07555430-3259589} & \colhead{2MASS~J07414279-0506464} & \colhead{2MASS~J19251275+0700362}
}
\startdata
RA (2MASS) [deg]                    &   118.976265     &   115.428332     &    291.303148 \\
DEC (2MASS) [deg]                   & $-$32.999706     &  $-$5.112907     &   $+$7.010075 \\
epoch (2MASS) [yr]                  &     1999.104     &     1998.866     &      1999.605 \\
$\mu_{\alpha}\cos{\delta}$ [mas/yr] & $-$120$\pm$7$^{\bullet}$     & $-$138$\pm$2$^{\blacktriangle}$     &   $+$46$\pm$4$^{\blacksquare}$ \\
$\mu_{\delta}$ [mas/yr]             & $+$161$\pm$3$^{\bullet}$     &  $+$31$\pm$2$^{\blacktriangle}$     &  $+$211$\pm$2$^{\blacksquare}$ \\
$l$ [deg]                           &     249.1868     &     223.3162     &       43.0435 \\
$b$ [deg]                           &    $-$2.3581     &    $+$8.7582     &     $-$4.2468 \\
$G$ (Gaia) [mag]                    &                  & (19.362$+$19.375)&        19.702 \\
$i$ (PS1) [mag]                     &                  & 19.021$\pm$0.011 & 19.986$\pm$0.062$^{\blacklozenge}$ \\
$z$ (PS1) [mag]                     &                  & 17.442$\pm$0.007 & 17.842$\pm$0.000$^{\blacklozenge}$ \\
$y$ (PS1) [mag]                     &                  & 16.453$\pm$0.006 & 16.850$\pm$0.017$^{\blacklozenge}$ \\
$J$ (2MASS) [mag]                   & 16.242$\pm$0.093 & 14.173$\pm$0.050 & 14.763$\pm$0.046 \\
$K_s$ (2MASS) [mag]                 & 13.527$\pm$0.041 & 12.389$\pm$0.034 & 12.937$\pm$0.037 \\
$W1$ (WISE all-sky) [mag]           & 12.473$\pm$0.025 & 11.635$\pm$0.023$^{\bigstar}$ & 12.048$\pm$0.023$^{\bigstar}$ \\
$W2$ (WISE all-sky) [mag]           & 11.866$\pm$0.024 & 11.313$\pm$0.022$^{\bigstar}$ & 11.661$\pm$0.022$^{\bigstar}$ \\
Spectral Type (photometric)         & $\approx$L7.5p   & $\approx$L5+L5   & $\approx$L7 \\
$d$ (photometric) [pc]              & $\approx$16      & $\approx$19      & $\approx$10 \\
\enddata
\tablecomments{$^{\bullet}$using all available epochs (2MASS, 8$\times$WISE), $^{\blacktriangle}$based on SSS, 2MASS, VHS, and mean position of two Gaia sources (excluded: 2$\times$DENIS, 8$\times$WISE, PS1), $^{\blacksquare}$based on 2MASS, first PS1 epoch, IPHAS, and Gaia (excluded: 7$\times$WISE, second PS1 epoch), $^{\blacklozenge}$available second epoch with similar photometry, $^{\bigstar}$merged with background object.}
\end{deluxetable}

\end{document}